# Observational Searches for Solar g-modes: Some Theoretical Considerations

Pawan Kumar[†], Eliot J. Quataert

Department of Physics

Massachusetts Institute of Technology, Cambridge, MA 02139

and

John N. Bahcall

Institute for Advanced Study, Princeton, NJ 08540

**Abstract:** We argue that the solar g-modes are unlikely to have caused the discrete peaks in the power spectrum of the solar wind flux observed by Thomson et al. (1995). The lower limit to the energy of individual g-modes, using the amplitudes given by Thomson et al., is estimated to be at least $10^{36}$ erg for low order g-modes; the resulting surface velocity amplitude is at least 50 cm s$^{-1}$, larger than the observational upper limit (5 cm s$^{-1}$).

We suggest that the most likely source for the excitation of solar g-modes is turbulent stresses in the convection zone. The surface velocity amplitude of low degree and low order g-modes resulting from this process is estimated to be of order $10^{-2}$ cm s$^{-1}$. This amplitude is interestingly close to the detection threshold of the SOHO satellite. The long lifetime of g-modes ($\sim 10^6$ years for low order modes) should be helpful in detecting these small amplitude pulsations.

*Subject headings:* Sun — oscillations: sun — g-modes

## 1. Introduction

Gravity mode oscillations of the sun are primarily confined to its radiative interior and their observation would thus provide a wealth of information about the energy generating region which is poorly probed by the p-modes. In the past 20 years a number of different groups have claimed to detect g-modes in the sun (e.g. Brookes et al. 1976; Brown et al. 1978; Delache & Scherrer 1983; Scherrer et al. 1979; Severny et al. 1976; for a detailed review of the observations please see the article by Pallé, 1991, and references therein), but thusfar there is no consensus that g-modes have in fact been observed. Recently Thomson, MacLennan & Lanzerotti (1995), hereafter referred to as TML, have reported detection of g-modes in the flux of high energy particles in the solar wind. In the absence of a detailed model of how pulsation affects the solar wind flux, it is difficult to determine the velocity amplitude of the pulsation in the photosphere from the amplitude of the solar wind flux variation. We, however, use general physical arguments to make an estimate of the photospheric velocity amplitude of g-modes from the wind data (section 2) which can be compared with the results of optical searches. In section 3 we provide estimates of the surface velocity amplitude, arising from excitation by turbulent convection, of g-modes of several different degree ($\ell$) and order ($n$) for comparison with this and future observations. The main results are summarized in section 4.

## 2. Observation of g-modes in the solar wind?

The dispersion relation for solar g-modes in the limit of $n \gg \ell$ is (e.g. Christensen-Dalsgaard & Berthomieu, 1991)

$$P_{n\ell} = \frac{P_0}{2\sqrt{\ell(\ell+1)}}(2n + \ell - \delta)$$

where

$$P_0 = \frac{2\pi^2}{\int_0^{r_1} dr\, N_B/r} \approx 2160 \text{ s},$$

$N_B$ is the Brunt-Väisälä frequency, $r_1$ is the upper turning point of the mode, and $\delta \approx 5/6$. Thus the order $n$ of a g-mode of frequency 5 $\mu$Hz (the low frequency end of the observed peaks attributed to g-modes by TML), and degree $\ell = 1$, 2, and 3 is about 130, 230, and 320, respectively. Thomson et al. find several groups of very closely spaced frequencies in their data, for instance 106.768, 111.938 and 114.042 $\mu$Hz (see Table 2 of TML). These can not be modes of the same $\ell$, if indeed they are g-modes, since the observed frequency spacing is considerably smaller than the frequency spacing expected for modes of fixed $\ell$, but larger than that due to rotational splitting (high degree g-modes have smaller frequency spacings, but they are unlikely to be observed because their surface velocity amplitudes are very small due to large attenuation in the convection zone). Thus, if g-mode amplitudes in the solar wind have a frequency dependence similar to that of table 2 of TML, then one expects there to be about 2000, $\ell$=1 & 2, g-modes with frequencies in the observed range, which should have amplitudes of order those detected by TML. (This is a conservative estimate since TML attribute some of their peaks to modes of $\ell > 2$.) Thomson et al., however, detect only about 40 such modes, which is a surprisingly small fraction of the total number of observable modes. Although there may be unstated reasons why other frequencies were not observed, it seems unlikely from an a priori point of view that such a small subset of frequencies should be detected.

The mean mass flux in the solar wind is $\sim 1.3 \times 10^{12}$ g s$^{-1}$ (or $2 \times 10^{-14} M_\odot$ per year), and the mean wind speed is 400 km s$^{-1}$. Thus the total kinetic energy flux in the wind is about $10^{27}$ erg s$^{-1}$; most of this energy is carried by the thermal particles, while the high energy particles observed by TML carry several orders of magnitude less energy. TML find the fractional variation in the high energy particle flux in the solar wind associated with g-mode pulsation to be about 2% for low frequencies ($\sim 10\mu$ Hz) and 0.3% for high frequencies ($\sim 100\mu$ Hz). Since the fractional variation in the magnetic field is also $\sim$2% (Thomson, personal communication), the flux of thermal particles in the wind is expected to be modulated by an amplitude similar to that of the high energy particles. Thus, provided that a significant fraction of the flux variation of the thermal particles is due to a change in the wind speed, TML's observations imply that the g-modes lose energy to the solar wind at a rate of about $10^{24}$ erg s$^{-1}$ for low $\ell$ modes with $n \gtrsim 10$ and $10^{22}$ erg s$^{-1}$ for modes with $n$ of order a few (the mean rate of energy loss is proportional to the square of the variation in the particle flux). An independent estimate of the rate of g-mode energy loss can be made using the observed variation of the magnetic field strength associated with a mode in the solar wind. These measurements are made using near-Earth satellites and are also reported by TML. Magnetic field variations are expected to propagate as Alfvén waves, which carry energy away from the sun at a rate of $\sim d^2 (\delta B)^2 V_A/2$, where $d$ is the distance of the spacecraft from the sun, $\delta B \sim 3 \times 10^{-6}$ gauss is the magnetic field variation associated with a mode (Thomson, personal communication), and $V_A \sim 50$ km s$^{-1}$ is the Alfvén speed in the solar wind. Thus, we infer that a g-mode loses approximately $5 \times 10^{21}$ erg s$^{-1}$ to Alfvén waves in the wind, which is consistent with the above estimate for low $n$ modes.

If we assume that the damping of the g-modes due to their modulation of the solar wind is not the dominant damping mechanism for these modes, then the lower limit to the g-mode energy is estimated to be about $10^{36}$ erg for low $\ell$ modes with $n$ of order a few and $10^{35}$ erg for low $\ell$ modes with $n \gtrsim 10$. These estimates were obtained by making use of the dissipation time (turbulent plus radiative) of g-modes, which is of order $10^6$ years for low order modes and decreases with $n$ as $n^2$ ($n$ is the mode order). The resultant surface velocity amplitude is at least 2 cm s$^{-1}$ for low $\ell$ modes with $n \gtrsim 10$ and 50 cm s$^{-1}$ for modes with $n$ of order a few. The latter amplitudes are larger than the observational upper limit of 5 cm s$^{-1}$ (Kuhn, Libbrecht and Dicke (1986) find a limit, for $\ell \geq 2$, of about 15 cm s$^{-1}$ at 100 $\mu$Hz and about 2 cm s$^{-1}$ at 10 $\mu$Hz; Garcia, Pallé and Cortés (1988) find a limit of about 4 cm s$^{-1}$ for low $\ell$, including $\ell$=1; Fröhlich (1990) also finds a limit of about 4 cm s$^{-1}$; for a review of the results of various optical searches for g-modes, see

Pallé, 1991, and references therein).

Thomson et al. also find p-modes in the solar wind electron flux data. The fractional variation of the electron flux associated with a 3 mHz p-mode is about $10^{-4}$ (Thomson, personal communication). Using the argument of the preceding paragraphs, with a mode lifetime of 10 days, we estimate the lower limit to the p-mode surface velocity amplitude to be about 1 cm s$^{-1}$; the observed amplitude is about 10 cm s$^{-1}$.

Particle flux perturbations imposed at the solar surface are dispersed out over a distance of a few tens of solar radii due to the distribution of particle speeds (this is damping due to free streaming of particles). Thus any periodic variation in the particle flux detected at a distance of $\sim$ 1 AU must arise due to in situ acceleration of particles. It is possible that g-modes excite Alfvén waves at the solar surface and that these waves in turn accelerate particles locally where they are observed. However, it is unclear if the observed magnetic field variation can lead to the variation of the high energy particle flux reported by TML. This is a complex problem which should be looked into carefully.

If we assume that perturbations associated with g-modes are imprinted on the solar wind magnetic field near the sun, which are carried away at the wind speed (away from the sun, the Alfvén speed is much less than the wind speed), then the phase shift observed in the variation of the magnetic field (and thus in the particle flux variation), at a distance $d$, over the course of a mode period, $P$, is $\sim 2\pi(d/PV_w)(\delta V_w/V_w)$, where $V_w$ is the mean thermal wind speed and $\delta V_w$ is the random variation of the wind speed over a time period $P$. Thus, the phase coherence time of a mode, observed in the solar wind, is estimated to be $\sim P(PV_w/d)^2(V_w/\delta V_w)^2$. The phase coherence time of a mode of period $\sim$ 2.6 hours (frequency 105$\mu$Hz), due to random variations in the wind speed of 1% in 2.6 hours, should therefore be of order a day. The coherence time observed by TML (fig. 2 of their paper) appears, however, to be about a year.

## 3. Velocity amplitude of g-modes due to turbulent excitation

A number of people have investigated the linear stability of solar g-modes (e.g. Dilke & Gough 1972; Rosenbluth & Bahcall 1973; Christensen-Dalsgaard et al. 1974; Shibahashi et al. 1975; Boury et al. 1975; Saio 1980). All of these investigations find that g-modes of radial-order ($n$) greater than 3 are stable. However, there is no general agreement about the stability of low order modes ($n \leq 3$). The reason for this is that there is a delicate balance between driving and damping for very low order modes, whereas higher order modes are stabilized by a rapid increase in damping with $n$. Thus it is almost certain that the high order g-modes reported by Thomson et al. can not be excited due to overstability.

Overstability, however, is not ruled out for very low order g-modes ($n \leq 3$ & $\nu \gtrsim$ 150$\mu$Hz). If overstable, the g-mode amplitude will increase exponentially with time until nonlinear effects become important and saturate their growth. Kumar and Goodman (1995) have recently investigated 3-mode parametric interaction, a very efficient nonlinear process. Using their results we find that the low order overstable g-modes in the sun will attain an energy of at least $10^{37}$ erg before they are limited by nonlinearities. The velocity at the solar surface corresponding to this energy is $\sim 10^2$ cm s$^{-1}$, which is an order of magnitude larger than the observational limit of Pallé (1991). Thus even low order g-modes of frequency greater than about 150 $\mu$Hz are unlikely to be overstable.

Stable g-modes can be excited by the same process that is responsible for exciting the p-modes, i.e. interaction with the turbulent convection. We follow the work of Goldreich, Murray and Kumar (1994) to estimate the energy and surface velocity amplitude of low degree g-modes. The excitation is due to the fluctuating Reynolds' stress and the theory has been calibrated so that it fits the energy input rate in the p-modes over the entire observed frequency range. We also estimate the turbulent and radiative damping of g-modes. These damping time scales are of the same order for modes of low $n$ ($\sim 10^6$ years), but the radiative damping rate increases with $n$ as $n^2$ whereas the turbulent damping rate is a weak function of mode order. Thus radiative damping dominates at $n$ greater than a few. Using the energy input rate and the damping rates we calculate the mode energy and the surface velocity amplitude for low degree g-modes.

Figure 1 shows the surface velocity amplitude as a function of frequency for low degree solar g-modes. Note that the surface velocity amplitude is about $10^{-2}$ cm s$^{-1}$ for low order g-modes and about $10^{-3}$ cm s$^{-1}$ for moderate to high order modes of $\ell$=1, which is a factor of about $10^4$ smaller than the amplitude inferred from Thomson et al.'s observation. We also find that the surface velocity amplitude falls off rapidly with $\ell$ making detection of high degree modes ($\ell \gtrsim 3$) extremely unlikely.

The ratio of the horizontal and radial velocity amplitudes, at the solar surface, can easily be shown to be equal to $[\ell(\ell+1)]^{1/2} GM_\odot/(R_\odot^3 \omega^2) \approx 0.13[\ell(\ell+1)]^{1/2} P_1^2$, where $P_1$ is the mode period in units of one hour. Thus, it is best to make Doppler measurements near the solar limb to search for low frequency g-modes.

## 4. Conclusion

Thomson et al. (1995) suggest that there are periodic variations in the high energy particle flux of the solar wind which they attribute to solar g-modes. They also suggest that g-modes are present in the solar magnetic field data. Since we expect there to be a variation of the thermal particle flux associated with the magnetic field variation, the oscillations should also be present in the flux of thermal particles, a testable prediction. Using the observed particle flux and magnetic field variations associated with g-modes, as reported by Thomson et al. (1995), we have estimated individual g-mode energies and find them to be at least $10^{36}$ erg (for comparison, the total energy in all of the solar p-modes is about $10^{33}$ erg). The corresponding surface velocity amplitude for a low order g-mode is at least 50 cm s$^{-1}$, which is larger than the observational upper limit (Pallé, 1991, and references therein) by about a factor of 10.

The phase shift experienced by a g-mode in the solar wind, as a result of a 1% variation in the thermal wind speed over the course of the mode period, should lead to an observed mode coherence time of order a day (for a 105$\mu$Hz mode), substantially shorter than that observed by TML (see §2).

High order solar g-modes ($n > 3$, or $\nu \lesssim 150\mu$Hz) are found to be stable by all of the published linear stability calculations. These g-modes can, however, be excited by turbulent stresses in the convection zone. The resulting velocity amplitude at the solar surface is estimated to be about $10^{-2}$ cm s$^{-1}$ for low order, low degree modes (our calculations of p-mode surface velocity amplitudes due to turbulent excitation fit the observed velocity spectrum well, and thus the error in the predicted g-mode velocity amplitudes is unlikely to be greater than an order of magnitude). This is about four orders of magnitude smaller than the amplitude implied by the observations of Thomson et al. The ground based network of telescopes designed to observe the solar oscillations continuously (GONG) may be able to detect g-mode pulsation down to about 1 cm s$^{-1}$, and SOHO is expected to have a detection threshold of about $10^{-1}$ cm s$^{-1}$ (Gabriel et. al, 1995). Thus g-modes, if stochastically excited, may be at the threshold of detectability of the forthcoming oscillation instruments. The expected long life time of g-modes ($\sim 10^6$ years for low order g-modes, and $10^3$ years for high order modes) should help in detecting these small amplitude pulsations.

**Acknowledgment:** We thank John Belcher, Tom Duvall, Peter Goldreich, Peter Sturrock, David Thomson, Louis Lanzerotti, and the referee for useful comments.

# REFERENCES

Boury, A., Gabriel, M., Noels, A., Scuflaire, R., & Ledoux, P. 1975, AA, 41, 279

Brookes, J.R., Isaak, G.R., and van der Raay, H.B. 1976, Nature 259, 92-95

Brown, T.M., Stebbins, R.T., and Hill, H.A. 1978, ApJ 223, 324

Christensen-Dalsgaard, J., and Berthomieu G. 1991, in *Solar Interior and Atmosphere*, eds. Cox, A.N., Livingston, W.C., and Matthews, M.S. (University of Arizona Press)

Christensen-Dalsgaard, J., Dilke, J. F. W. W., & Gough, D. O. 1974, M.N.R.A.S., 169, 429

Delache, P., and Scherrer, P.H. 1983, Nature 306, 651

Dilke, J. F. W. W., & Gough, D. O. 1972, Nature, 240, 262

Dziembowski, W.A. 1983, Solar Phys. 82, 259-266

Fröhlich, C. 1990, in *Oji Seminar on Progress of Seismology of the Sun and Stars – Hakone, Japan, December 11-14, 1989, Lecture Notes in Physics*, ed. H. Shibahashi, Springer-Verlag, New York

Gabriel, A. et. al 1995, to appear in Solar Phys

Garcia, C., Pallé, and Cortés, T. Toca 1988, in *Seismology of the Sun and Sun-like Stars*, ed. E.J.Rolfe, ESA SP-286, 353

Goldreich, P., Murray, N., and Kumar, P. 1994, ApJ 424, 466

Kuhn, J.R., Libbrecht, K.G., and Dicke, R.H. 1986, Nature 319, 128-131

Kumar, P., and Goodman, J. 1995, ApJ

Pallé, P. L. 1991, Adv. Space Res., 11, 4, 29

Rosenbluth, M., & Bahcall, J. N. 1973, ApJ, 184, 9

Saio, H. 1980, ApJ, 240, 685

Scherrer, P.H., Wilcox, J.M., Kotov, V.A., Severny, A.B., and Tsap, T.T. 1979, Nature 277, 635

Severny, A.B., Kotov, V.A., and Tsap, T.T. 1976, Nature 259, 87

Shibahashi, H., Osaki, Y., & Unno, W. 1975, Publ. Astron. Soc. Japan, 27, 401

Thomson, David J., Maclennan, Carol G., Lanzerotti, Louis J., 1995, Nature, 376, 139 (TML)

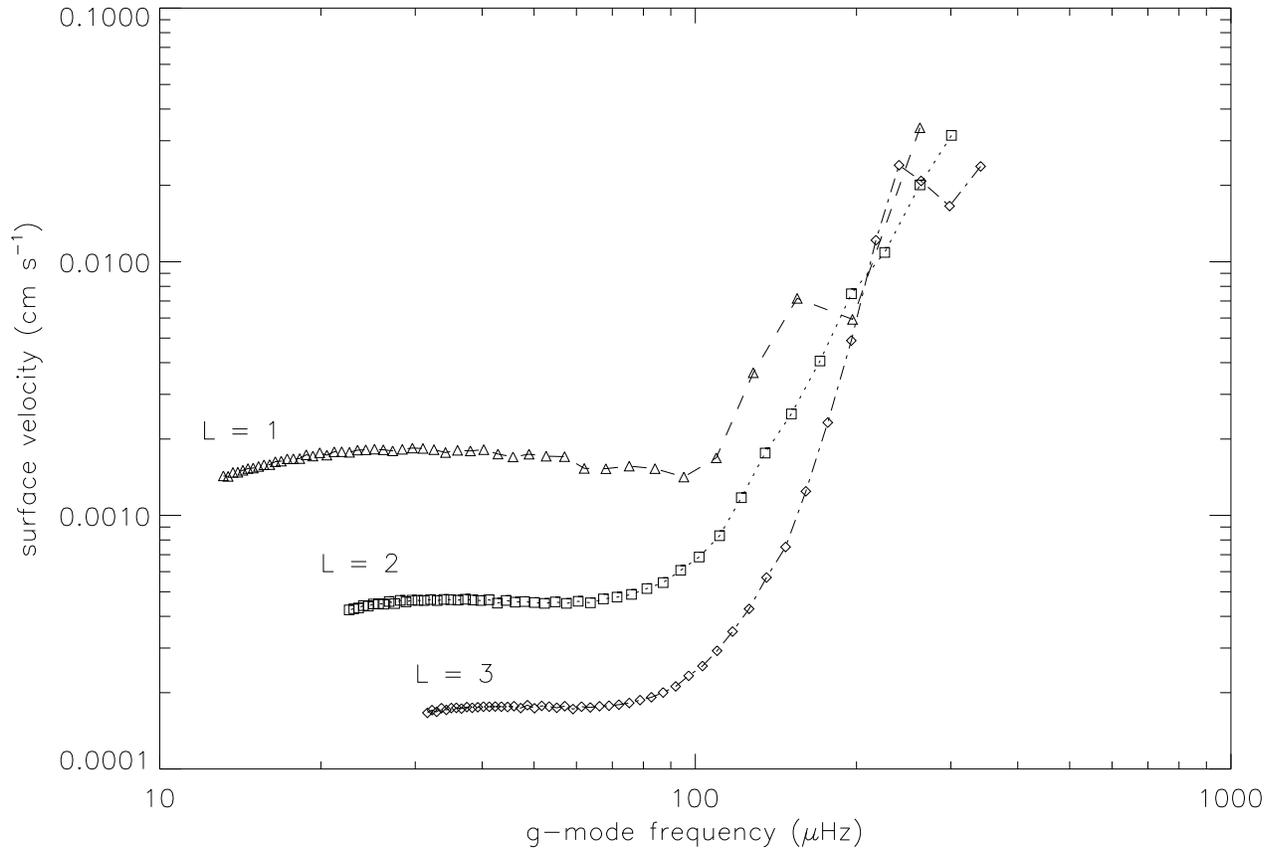

FIG. 1.—Magnitude of the surface velocity amplitude as a function of frequency for low degree solar g-modes excited by coupling with turbulent convection. The surface velocity amplitude falls off strongly with increasing $\ell$ making detection of high $\ell$ modes unlikely.

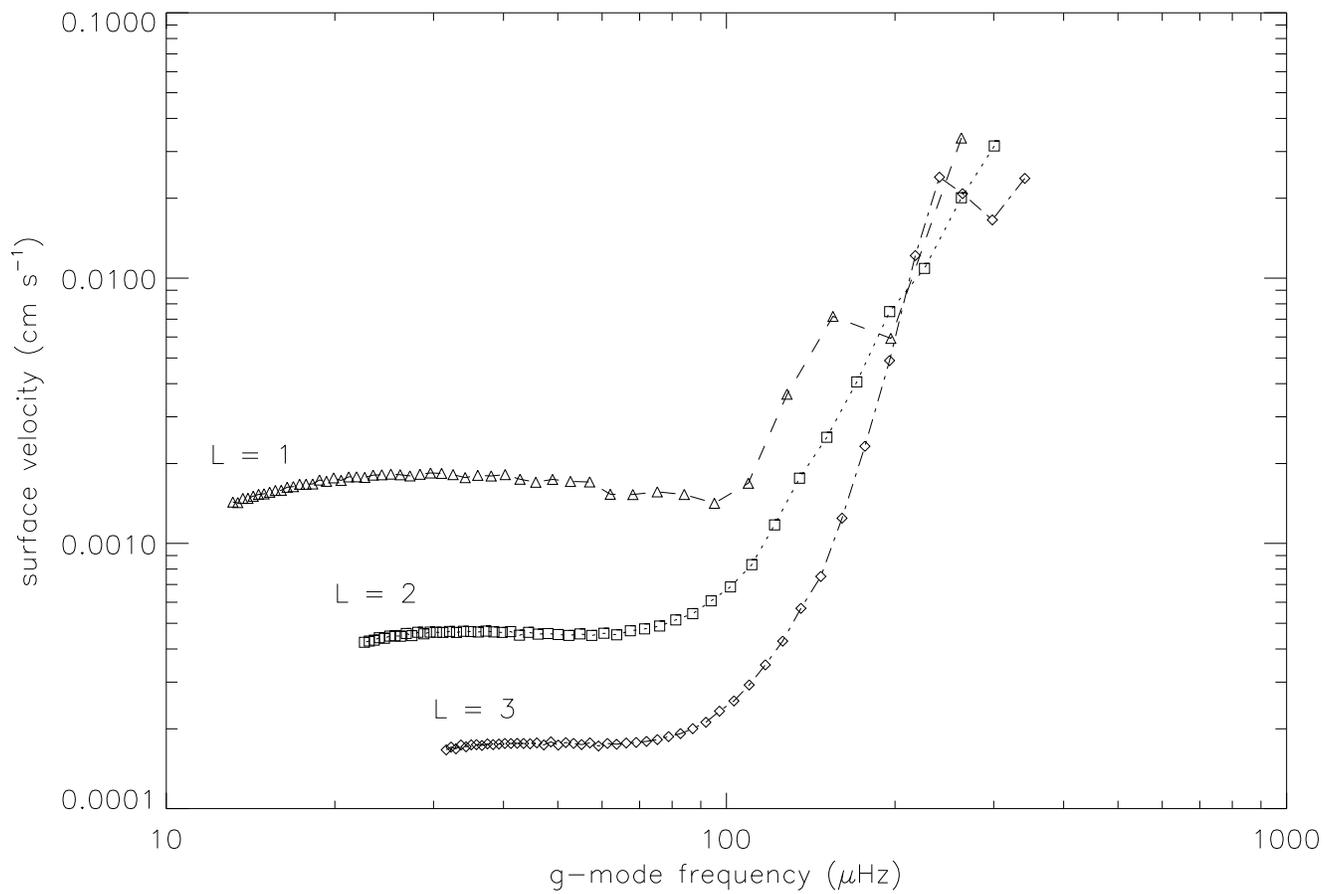